\newcommand{\NDMAP}{Ni(C$_5$D$_{14}$N$_2$)$_2$N$_3$(PF$_6$)}
\newcommand{\NENP}{Ni(C$_2$H$_8$N$_2$)$_2$NO$_2$(ClO$_4$)}
\newcommand{\NDMAZ}{Ni(C$_5$H$_{14}$N$_2$)$_2$N$_3$(ClO$_4$)}

\documentclass[aps,prb,twocolumn,superscriptaddress,showpacs]{revtex4}
\usepackage{graphicx}
\usepackage{natbib}
\usepackage{bm}

\begin{document}

\title{Dynamics of an anisotropic Haldane antiferromagnet in strong magnetic field}

\author{A. Zheludev}
\affiliation{Condensed Matter Sciences Division, Oak Ridge
National Laboratory, Oak Ridge, TN 37831-6393, USA.}
 \email{zheludevai@ornl.gov}
 \homepage{http://neutron.ornl.gov/~zhelud/}

\author{S. M. Shapiro}
\affiliation{Physics Department, Brookhaven National Laboratory,
Upton, NY 11973-5000, USA.}

\author{Z. Honda}
\affiliation{Faculty of Engineering, Saitama University, Urawa,
Saitama 338-8570, Japan.}

\author{K. Katsumata}
\affiliation{The RIKEN Harima Institute, Mikazuki, Sayo, Hyogo
679-5148, Japan.}

\author{B. Grenier}
\author{E. Ressouche}
\author{L.-P.~Regnault}
\affiliation{DRFMC/SPSMS/MDN, CEA-Grenoble, 17 rue des Martyrs,
38054 Grenoble Cedex, France.}

\author{Y. Chen}
 \affiliation{Department of Physics and Astronomy, Johns Hopkins
University, Baltimore, MD 21218, USA.}
 \affiliation{Present
address: Los Alamos Natl. Laboratory, Los Alamos, NM 87545, USA.}

\author{P. Vorderwisch}
\affiliation{BENSC, Hahn-Meitner Institut, 14109 Berlin, Germany.}

\author{H.-J. Mikeska}
\affiliation{Institut f\"ur Theoretische Physik, Universit\"at
Hannover, Appelstra{\ss}e 2, 30167 Hannover, Germany.}

\author{A. K. Kolezhuk}
\affiliation{Institut f\"ur Theoretische Physik, Universit\"at
Hannover, Appelstra{\ss}e 2, 30167 Hannover, Germany.}
\affiliation{Institute of Magnetism, National Academy of Sciences and
Ministry of Education, 36(b) Vernadskii av., 03142 Kiev, Ukraine }

\date{\today}
\begin{abstract}

We report the results of elastic and inelastic neutron scattering
experiments on the Haldane-gap quantum antiferromagnet \NDMAP\
performed at mK temperatures in a wide range of magnetic field
applied parallel to the $S=1$ spin chains. Even though this
geometry is closest to an ideal axially symmetric configuration,
the Haldane gap closes at the critical field $H_{c}\simeq 4T$, but
reopens again at higher fields.  The field dependence of the two
lowest magnon modes is experimentally studied and the results are
compared with the predictions of several theoretical models.  We
conclude that of several existing theories, only the recently
 proposed model [Zheludev et al., cond-mat/0301424 ] is able
to reproduce all the features observed experimentally for different
field orientations.

\end{abstract}


\maketitle

\section{Introduction}

The problem of field-induced magnon condensation in gapped quantum
antiferromagnets is currently receiving a great deal of attention
from experimentalists. Particularly important results were
obtained in recent neutron scattering and ESR measurements on
$S=1$ Haldane-gap\cite{Haldane83,Haldane83-2} compounds \NENP\
(NENP),\cite{Enderle00} \NDMAP\ (NDMAP)
\cite{Honda98,Honda99,Chen01,Zheludev2002,Zheludev+03,Hagiwara+03}
and \NDMAZ\ (NDMAZ),\cite{Zheludev2001-2}  as well as the
$S=\frac{1}{2}$-dimer system
TlCuCl$_3$.\cite{Nikuni+00,Ruegg+02,Ruegg+03} The effect of
magnetic field is to drive the gap $\Delta$ in such systems to
zero by virtue of Zeeman effect, thus promoting a quantum phase
transition to a new magnetized state at some critical value of the
applied field $H_c\approx (\Delta/g\mu_{B})$.\cite{Katsumata+89}
Additional magnetic anisotropy effects usually lead to more
complex behavior and richer phase diagram. Anisotropy is
negligible in many $S=\frac{1}{2}$-based materials such as
TlCuCl$_3$, where no single-ion terms are possible. In contrast,
for $S=1$ compounds such as NENP, NDMAP and NDMAZ, terms of type
$DS_z^2$ are quite strong and the corresponding zero-field
anisotropy splitting of the excitation triplet is comparable in
magnitude to the Haldane gap itself.  Under these circumstances,
the physics is expected to depend strongly on the direction of the
applied magnetic field with respect to the anisotropy
axes.\cite{Golinelli93-2}

For purely technical reasons, in quasi-1D materials it is much
easier to perform inelastic neutron scattering experiments in high
magnetic fields applied \emph{perpendicular} to the spin chains. On
most instruments the scattering plane is horizontal and the wave
vector resolution along the vertical axis is deliberately
coarsened to provide an intensity gain. To optimize wave vector
resolution along the spin chains in the sample, and to allow the
momentum transfer $q_\parallel\approx\pi$ in that direction, the chain
axis has to be mounted in the horizontal plane. The typical
construction of superconducting magnets is such that the field is
along the vertical direction, and is thus applied perpendicular to
the spin chains in the sample. In the Haldane-gap materials NENP,
NDMAP and NDMAZ the anisotropy easy plane is roughly perpendicular
the chains. As a result, most of the previous neutron measurements
were performed for \emph{in-plane} magnetic fields, i.e., in the
Axially Asymmetric (AA) geometry. In the AA case the transition at
$H_c$ is expected to be of Ising type,\cite{Tsvelik90,Affleck91} and
even an isolated chain acquires antiferromagnetic long-range order
in the ground state at $H>H_c$. The excitations in the magnetized
state are a triplet of massive ``breathers'' (soliton-antisoliton
bound states).\cite{Affleck91} Recent neutron scattering studies of NDMAP in the
AA geometry provided a solid confirmation of these theoretical
predictions.\cite{Zheludev+03}

For an isolated Haldane spin chain magnetized in the Axially
Symmetric (AS) geometry (with a magnetic field applied parallel to
the anisotropy axis and the anisotropy being of a purely
easy-plane type) theory predicts a totally different, disordered
and quantum-critical ground state whose low-energy physics can be
described as the Tomonaga-Luttinger spin
liquid.\cite{Affleck91,Tsvelik90,Takahashi,Sachdev+94} The
low-energy excitation spectrum contains no sharp modes and is
instead a continuum of states, much as for $S=\frac{1}{2}$ spin
chains.\cite{Schulz+83-86} Higher modes which would have
quasiparticle character in the AA geometry (two upper members of
the Zeeman-split triplet) also develop into continua and exhibit
only edge-type singularities in the AS
case.\cite{KolezhukMikeska02} Moreover, one expects incommensurate
correlations at a field-dependent Fermi wave vector that
characterizes the Fermi sea of magnons ``condensed''  at
$H>H_c$.\cite{Furusaki99} In a real material this idealized
picture may be complicated by several factors. First of all, there
can be additional anisotropy terms, such as in-plane single-ion
anisotropy or Dzyaloshinskii-Moriya interactions, which
\emph{explicitly} break the axial symmetry and favor the AA
physics; a similar effect could be expected if the direction of
the applied field deviates slightly from the symmetry axis.
Residual three-dimensional inter-chain couplings lead to a
\emph{spontaneous} breaking of the axial symmetry which is
equivalent to the Bose-Einstein condensation of
magnons.\cite{Affleck91,Sachdev+94,GiamarchiTsvelik99,Nikuni+00}
 Due to the critical nature of the ground state of an isolated
chain, all these effects are relevant and will have a significant
impact on the spin correlations, no matter how small they may be.
The key to understanding the high-field behavior in real materials
is a combined approach involving high resolution neutron
scattering experiments and a consistent theoretical treatment. Due
to geometrical constraints described in the previous paragraph,
high-field neutron measurements in the AS configuration are
technically challenging and require the use of specialized neutron
instruments or magnet environments.

The purpose of the present paper is twofold: on one hand, to study
experimentally the response of the NDMAP system in the wide range
of applied fields and at very low temperature, in the setup as
close to the AS geometry as possible, and to test further the
phenomenological field theory of the high-field phase which was
recently proposed \cite{Zheludev+03} and successfully applied to
the description of experiments on NDMAP in the AA geometry.
\cite{Hagiwara+03}  On the other hand, we also give a more
detailed account of the theory which was only briefly outlined in
Ref.~\onlinecite{Zheludev+03}, and generalize it to include the
effect of interchain interactions. Finally, we perform a
systematic quantitative comparison between our experimental
findings and theoretical results.

The paper is organized as follows: in Sect.\ \ref{sec:exp} we
present the details of experimental setup, and Sect.\
\ref{sec:results} reports the results of elastic and inelastic
neutron scattering measurements. In Sect.\ \ref{sec:theory} we
present the general effective field theory of anisotropic gapped
quasi-one-dimensional spin system in strong magnetic field and
perform a systematic quantitative comparison between our
experimental findings and theoretical results. Finally, Sect.\
\ref{sec:summary} contains the summary and concluding remarks.

\section{Experimental setup}
\label{sec:exp}

\subsection{Structural considerations and energy scales}

\begin{figure}
\includegraphics[width=3.3in]{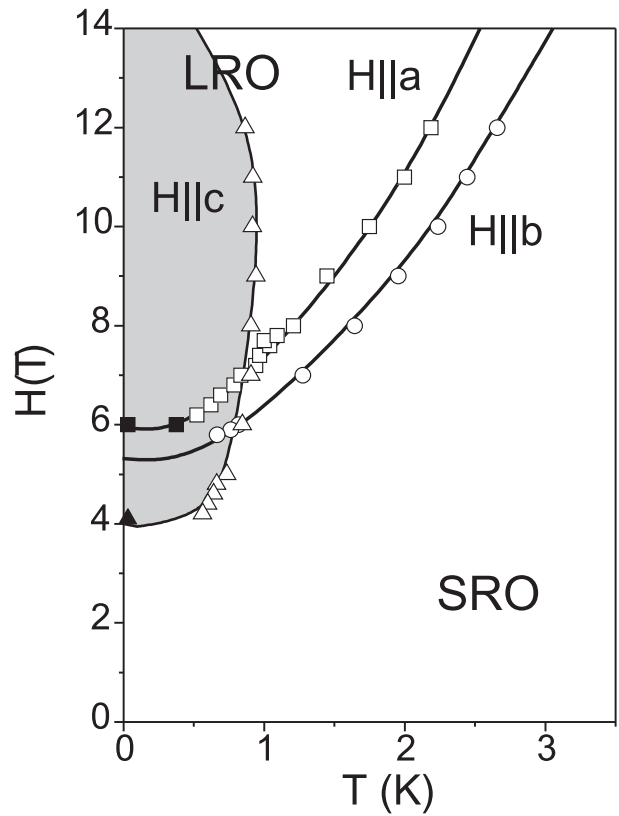}
 \caption{\label{phase} The $H-T$ phase diagram of NDMAP as deduced from
 specific heat measurements (open symbols, Ref.~\protect\onlinecite{Honda98}) and
 neutron scattering experiments (solid symbols, Refs.~\protect\onlinecite{Chen01,Zheludev+03} and
 the present study). The shaded area is the high-field phase investigated
 in the present work. }
\end{figure}

The crystal structure of NDMAP is schematically shown in Fig.~1 of
Ref.~\onlinecite{Zheludev2001}. The AF spin chains are composed of
octahedrally-coordinated $S=1$ Ni$^{2+}$ ions bridged by
azido-groups. The chains run along the $c$ axis of the
orthorhombic structure (space group $P {nmn}$, $a=18.046$~\AA,
$b=8.705$~\AA, and $c=6.139$~\AA). Previous zero-field neutron
studies provided reliable estimates for the relevant magnetic
energy scales in the system. The in-chain exchange constant is
$J=2.6$~meV. Exchange coupling along the crystallographic $b$ axis
is considerably weaker, $J_b/J\approx 10^{-3}$, and that along $a$
is weaker still, to the point of being undetectable:
$|J_a/J|<10^{-4}$. Magnetic anisotropy in NDMAP is predominantly
of single-ion easy-plane type with $D/J\approx 0.25$. In addition,
there is a weak in-plane anisotropy term of type
$E(S_{x}^2-S_{y}^2)$. As a result of these anisotropy effects, at
zero field the degeneracy of Haldane triplet is fully lifted and
the gap energies are $\Delta_1=0.42(3)$~meV,
$\Delta_2=0.52(6)$~meV, and $\Delta_3=1.9(1)$~meV.
Correspondingly, the critical fields are strongly dependent on
field orientation and the $H-T$ phase diagram, visualized in
Fig.~\ref{phase} is highly anisotropic.

\begin{figure}
\includegraphics[width=3.3in]{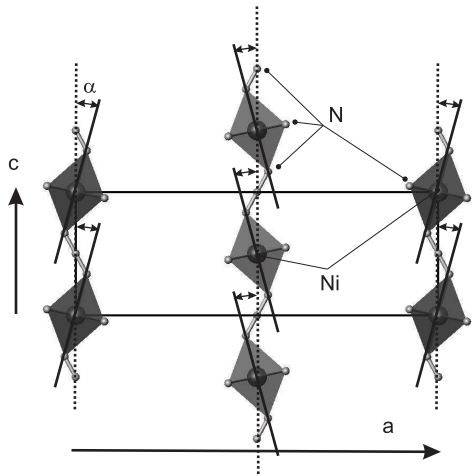}
 \caption{\label{str} Main elements of the NDMAP crystal structure in projection onto
 the $(ac)$ crystallographic plane. Only the N and Ni sites are shown. Magnetic
 anisotropy associated with the Ni$^{2+}$ magnetic ions is determined by the local
 symmetry of NiN$_6$ coordination tetrahedra that are tilted in the $(ac)$ plane by
 $\alpha=16^{\circ}$ relative to the $c$ axis. The tilt direction is opposite in
 adjacent chains.}
\end{figure}

The local anisotropy axes in NDMAP are determined by the geometry
of the corresponding Ni$^{2+}$ coordination octahedra and do {\it
not} exactly coincide with crystallographic directions. Instead,
the main axes of the NiN$_6$ octahedra are in the $(a,c)$
crystallographic plane, but tilted by  $\alpha\approx 16^\circ$
relative to the $c$ axis. As illustrated in Fig.~\ref{str}, within
each chain, the tilts are in the same direction for all
Ni$^{2+}$-sites, so there is no intrinsic alternation in the
chains, as is the case in many related compounds such as NENP.
However, within the crystal structure there are two types of
chains related by symmetry, and the corresponding tilt directions
are opposite. This circumstance has a very important consequence
for the present study. It implies that for NDMAP {\it one can not
apply the field along the anisotropy axis of all the magnetic ions
in the sample}. The closest one can come to this idealized AS
scenario is by applying a field along the axis of {\it bulk}
magnetic anisotropy, i.e., along the $(0,0,1)$ direction. In this
case the field will form a small angle of $\pm \alpha$ with the
{\it local} anisotropy axes for all spin chains.

\subsection{Experimental procedures}

Single-crystal neutron scattering experiments in magnetic fields
applied parallel to the spin chains were carried out using two
different setups. A series of diffraction measurements was carried
out using a vertical-field 6~Tesla cryomagnet installed on the
rather unique D23 lifting counter diffractometer at ILL. On this
instrument the data collection is not restricted to a given
scattering plane. The sample is mounted with the chain axis
vertical (parallel to the field) and the detector is lifted out of
the horizontal plane to allow a momentum transfers
$q_\parallel\approx\pi$ in that direction. Neutrons of a
fixed-incident energy $E_i=14.7$~mev were provided by a curved
thermal neutron guide with supermirror coating, a Pyrolitic
Graphite (PG) monochromator and a PG filter. In some cases
horizontal and/or vertical collimators with a 40' FWHM beam
acceptance were inserted in front of the detector. In the
diffraction study we employed a $\approx 0.3$~g fully deuterated
single crystal sample.

Inelastic measurements were performed using the conventional
3-axis cold neutron spectrometer FLEX installed at HMI. An
assembly of 3 deuterated single crystals with total mass about 1~g
were mounted with chain axis in the horizontal (scattering) plane
of the instrument. A magnetic field was applied along that
direction by a horizontal-field cryomagnet. The data were
collected with the final neutron energy fixed at 5~meV. A Be
filter was used after the sample to eliminate higher-order beam
contamination. Beam divergencies were defined by the critical
angle of the cold-neutron guide, and by characteristics of the PG
analyzer and monochromator used. No additional devices were used
to collimate the neutron beams. In both the D23 and FLEX
experiments the sample environment was a $^{3}$He-$^{4}$He
dilution refrigerator.

\section{Results of neutron scattering measurements}
\label{sec:results}

\subsection{Magnetic long range order}

\begin{figure}
\includegraphics[width=3.3in]{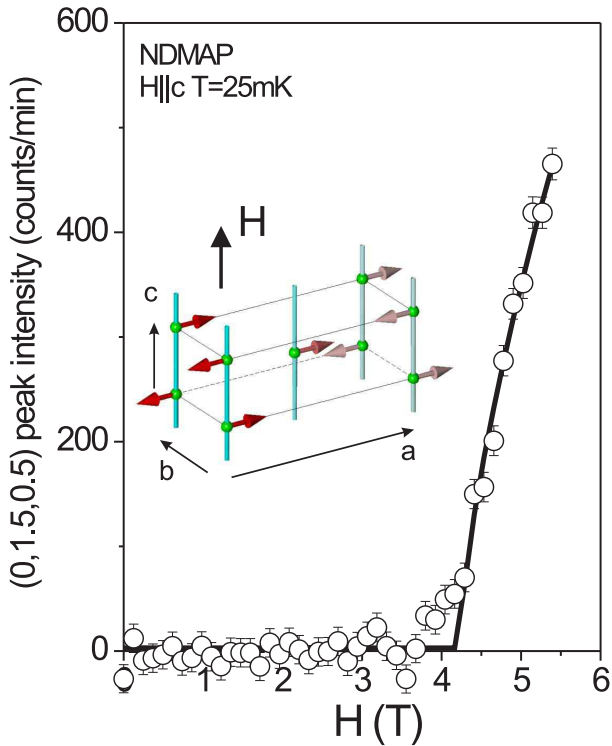}
\caption{\label{Bvsh}
 Measured peak intensity of the $(0,1.5,0.5)$ magnetic Bragg reflection in NDMAP as a
 function of magnetic field applied along the crystallographic $c$ axis
 (symbols). The solid line is a power-law fit to the data, that should only be viewed
 as a guide for the eye. Inset: proposed model for the spin structure in
 the high-field phase. Only the staggered magnetization probed by neutron diffraction
 is shown. In addition, a weak ``ferromagnetic'' tilt of all spins along the field direction is expected.}
\end{figure}

Applying a field $H>H_c\approx 4.0$~T parallel to the $c$ axis at
$T=25$~mK leads to antiferromagnetic long-range ordering in NDMAP.
This was deduced from the appearance of new Bragg reflections at
the 3D AF zone-centers $(h,k,l)+(0,-0.5,0.5)$, where $h$, $k$ and
$l$ are integer. The measured field dependence of the (0,1.5,0.5)
background-subtracted peak intensity is plotted in
Fig.~\ref{Bvsh}. In this data set the experimental error bars are
too large to allow an accurate determination of $T_{N}$ and the
order-parameter critical index.

\begin{figure}
 \includegraphics[width=3.3in]{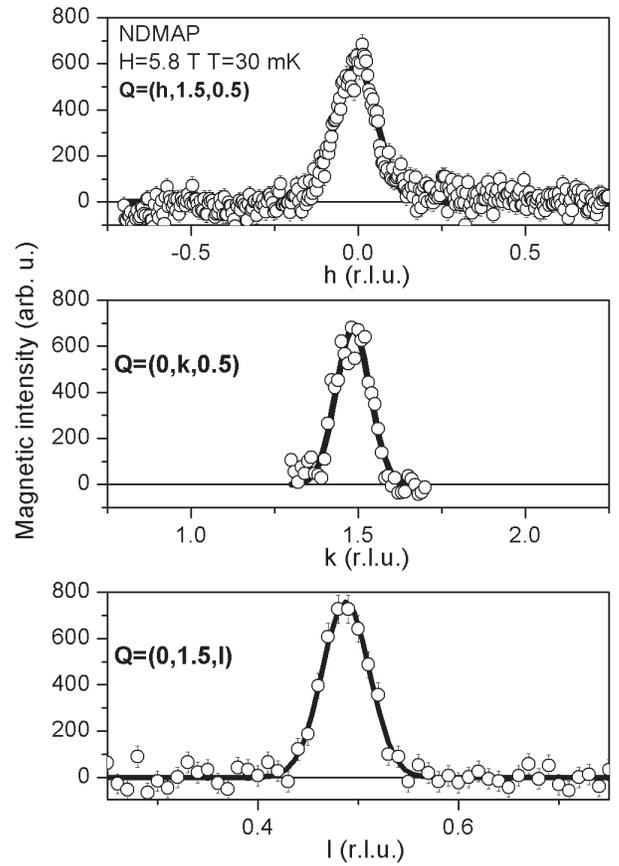}
 \caption{\label{diffraction} Elastic scans across the $(0,1.5,0.5)$ magnetic Bragg
 reflection in NDMAP at $T=30$~mK and $H=5.8$~T show resolution-limited peaks in all
 three directions.  Solid lines are Gaussian fits to the data.}
\end{figure}

Fig.~\ref{diffraction} shows $h$, $k$ and $l$ scans across the
$(0,1.5,0.5)$ peak measured at $H=5.8$~T. These data were taken
with a horizontal collimator installed in front of the detector,
to improve wave vector resolution along the chain axis. The
background was separately measured at $H=0$ and subtracted from
the data shown. In all scans the magnetic peak width was
determined to be resolution-limited. Thus, to within the
experimental wave vector resolution, the high-field phase in NDMAP
for $H\parallel c$ is characterized by true 3-dimensional
long-range order.

To determine the high-field spin arrangement, integrated
intensities of 17 magnetic Bragg reflections were measured at
$T=25$~mK and $H=5.8$~T using a standard diffraction configuration
(no collimators). The observed intensity pattern is well
reproduced by the simple model for the magnetic structure
illustrated in the insert of Fig.~\ref{Bvsh}. This  spin
arrangement is the same as previously seen for $\mathbf{H}\|[0
\overline{1}, 1]$.\cite{Chen01} The collinear model is clearly
oversimplified, and the actual structure should be a canted one,
with all spins slightly tilted towards the field and thus
producing a net magnetization along that direction. In our
experiment we can detect only the staggered part $\vec{L}$ of the
magnetic moment which lies, within the experimental accuracy,
along the crystallographic $a$ axis, perpendicular to the field
direction. The local spin directions are opposite on sites related
by $(0,1,0)$ and $(0,0,1)$ translations, and the same on sites
related by a translation along $(1,0,0)$. Comparing the
experimentally determined intensities of nuclear and magnetic
reflections provides an estimate for the total staggered
magnetization per site: $L=0.9(1)$~$\mu_{\mathrm{B}}$. This
experimental value is about half of the classical sublattice
magnetization for an ordered $S=1$ system.

\subsection{Inelastic scattering}

\subsubsection{Constant-$q$ data}

\begin{figure}
\includegraphics[width=3.5in]{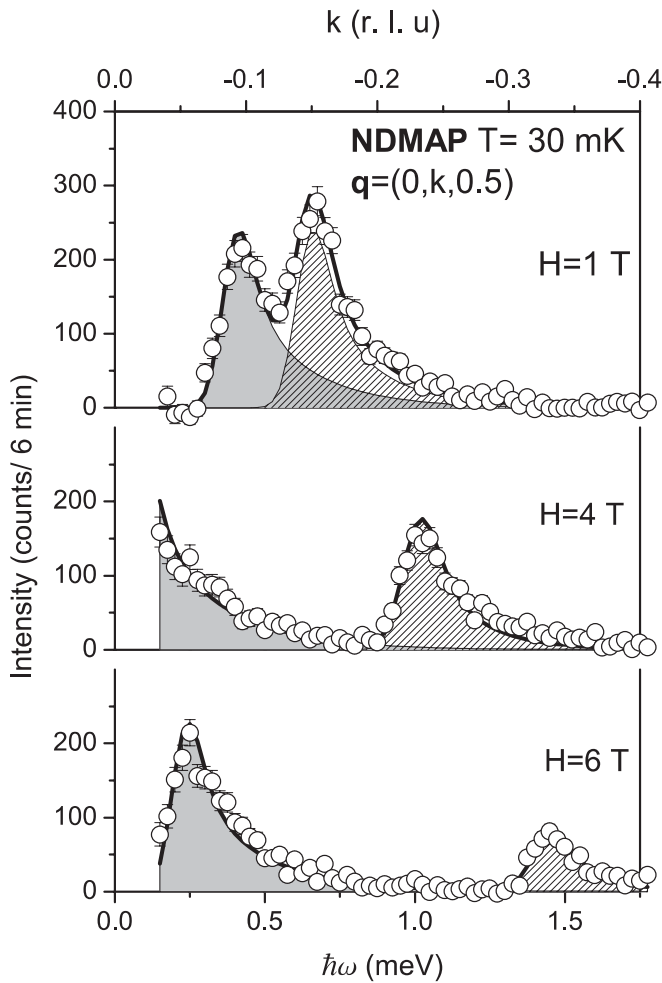}
 \caption{\label{exdata1} Typical background-subtracted
 constant-$q_\parallel$ scans measured in NDMAP up to 1.75~meV energy
 transfer for different values of magnetic field applied along the
 crystallographic $c$ axis (symbols). The two peaks are the
 lower-energy members of the Haldane excitation triplet. The third
 mode has a larger gap and is outside the shown scan range. Solid
 lines are fits to the data using a single-mode model cross
 section, as described in the text.}
\end{figure}

The field dependence of the gap energies was measured in a series
of energy scans performed at the 1D AF zone-center $(0,k,0.5)$.
The scans corresponded to a fixed momentum transfer parallel to
the chains $q_\parallel=\pi/c$. The wave vector transfer
perpendicular to the chains $q_\perp=2\pi k/b$ was varied in the
course of the scan to satisfy geometrical restrictions imposed by
the construction of the horizontal-field magnet.  The background
for all scans was measured at $H=0$ and $H=6$~T, away from the 1D
AF zone-center, at $\mathbf{q}=(0,k,0.33)$. Apart from the
expected elastic contribution due to incoherent scattering, the
background was found to be energy-independent and about 1.5
counts/min.  Typical background-removed scans collected at
$H=1$~T$<H_c$, $H=4.1$~T$\approx H_c$ and $H=6$~T$>H_c$ are shown
in Fig.~\ref{exdata1} (symbols).

\begin{figure*}
\includegraphics[width=6in]{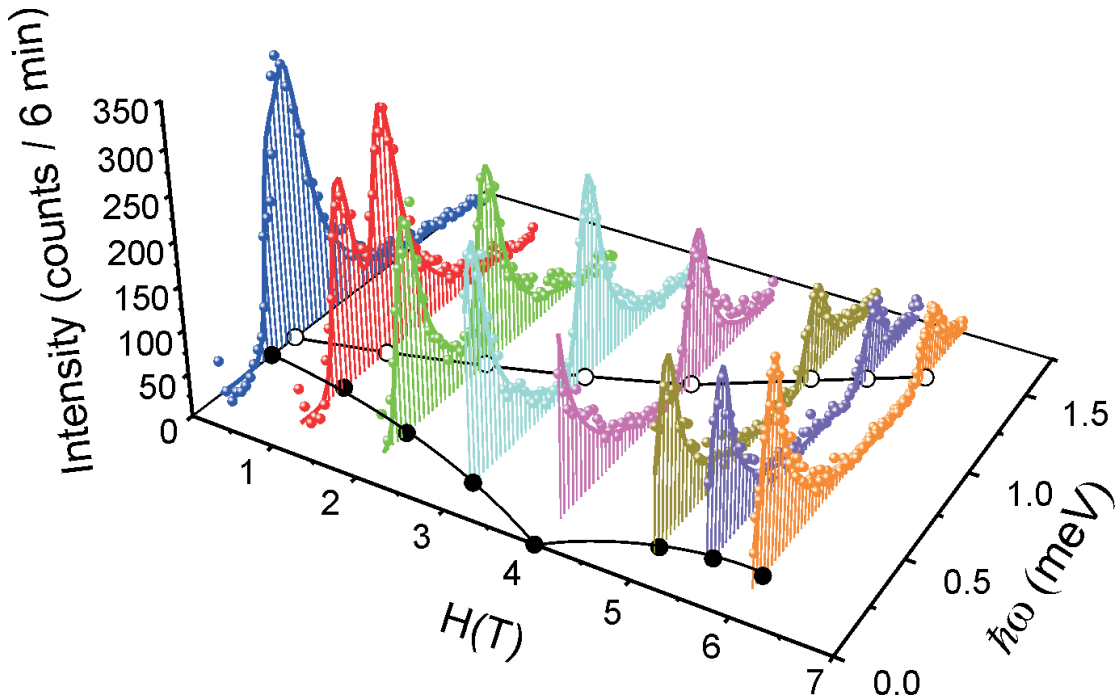}
 \caption{\label{exdata2} Background-subtracted constant-$q_\parallel$
 scans measured in NDMAP for different values of magnetic field
 applied along the crystallographic $c$ axis. Solid lines and
 shaded areas are fits to the data using a model cross section
 function, as described in the text. The black and white circles
 in the $(H,\hbar\omega)$ plane are the measured field dependence
 of the two lower energy gaps in NDMAP. The connecting solid lines
 are guides for the eye.}
\end{figure*}

Scans collected at all fields are combined in the 3D plot shown in
Fig.~\ref{exdata2}.  The data were analyzed using a single-mode
cross section function similar to that used in
Ref.~\onlinecite{Zheludev2001}: \begin{eqnarray}
S_j(\mathbf{q},\omega) & \propto &
|f(q)|^2P_j\frac{1-\cos(\mathbf{q}\mathbf{c})}{2\omega_{j,\mathbf{q}}}\times\nonumber\\
& \times & \left[\delta(\hbar \omega-\hbar \omega_{j,\mathbf{q}})+\delta(\hbar
\omega+\hbar \omega_{j,\mathbf{q}})\right]\\ (\hbar \omega_{j,\mathbf{q}})^2 & = &
\Delta_j^2+v^2\sin^2(\mathbf{q}\mathbf{c}).  \end{eqnarray} Here $j={1,2,3}$ labels
each of the three excitation branches, $v$ is the spin wave velocity, $f(q)$ is the
magnetic form factor for Ni$^{2+}$, and $\Delta_j$ are the gap energies. The
intensity prefactors $P_j$ depend on both the matrix elements between the ground
state and the single-mode excited states and on the polarization of the latter. The
model cross section was numerically convoluted with the calculated spectrometer
resolution function. The gap energies and prefactors for each mode were refined using
a least-squares routine to best-fit the data collected at each field. The spin wave
velocity was fixed at $v=6.5$~meV, as determined previously for
$H=0$.\cite{Zheludev+03} The resulting fits are represented by the solid lines in
Figs.~\ref{exdata1} and \ref{exdata2}. The shaded areas in Fig.~\ref{exdata1} are
partial the contributions of each mode. The obtained field dependence of the energy
gap is plotted in the $(H,\hbar\omega)$ plane of the 3D plot in Fig.~\ref{exdata2}
and, in more detail, in the top panel of Fig.~\ref{gapvsH}. We have observed only two
lower-energy members of the Haldane excitation triplet; the third mode has a larger
gap and is outside the shown scan range. The field dependence of the upper two
triplet modes was observed in recent ESR measurements.\cite{Hagiwara+03} However, the
lowest triplet mode was not observed in Ref.\ \onlinecite{Hagiwara+03} since those
measurements were done at much higher temperature $T=1.5$~K where the lowest mode
becomes strongly damped, the situation similar to that encountered in our early
experiments in the $H\parallel a$ geometry which were also done at high temperature
and failed to observe the reopening of the gap in the lowest mode.

\begin{figure}
\includegraphics[width=3.5in]{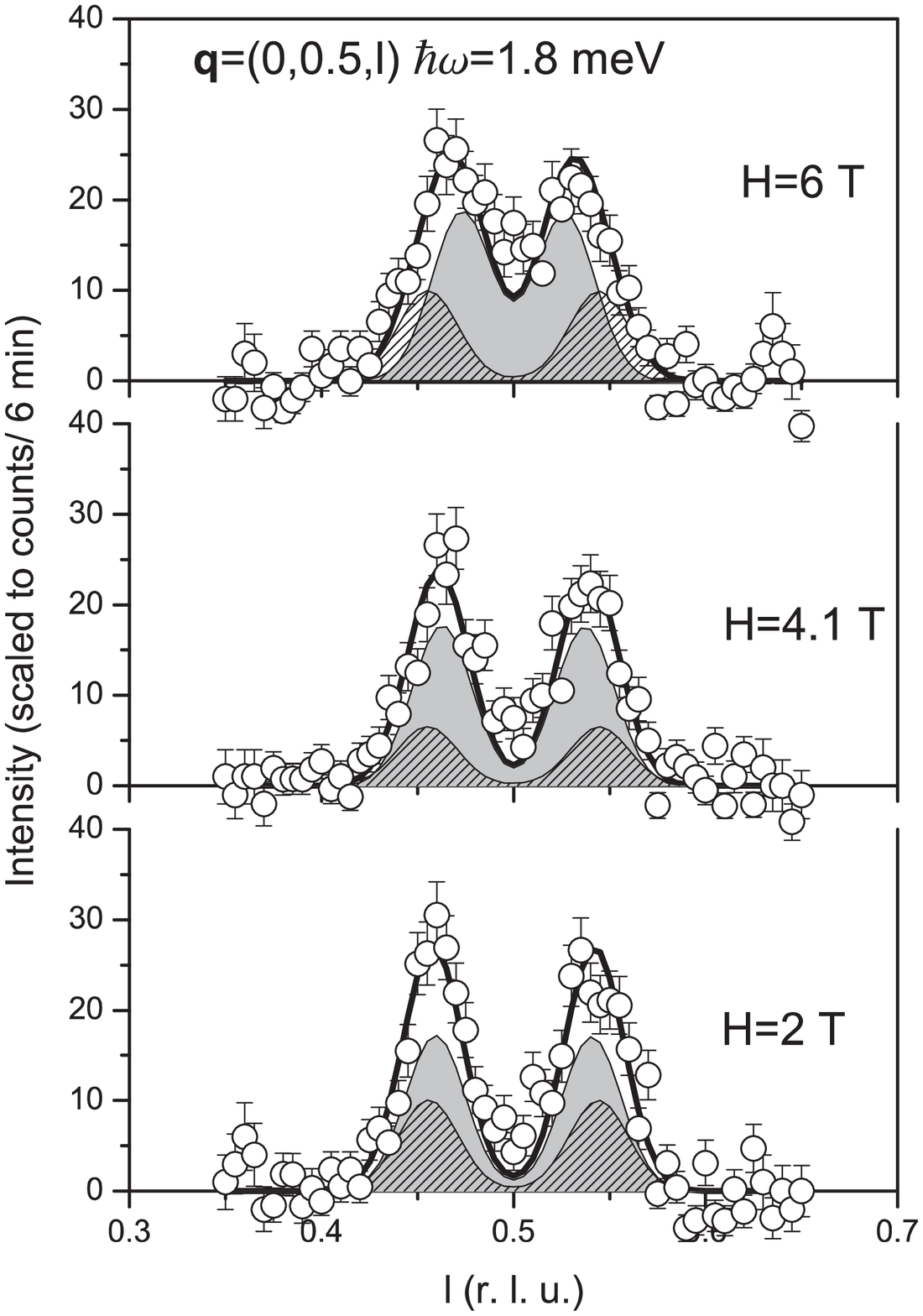}
 \caption{\label{exdata3} Background-subtracted constant-$E$ scans
 measured in NDMAP for different values of magnetic field applied
 along the crystallographic $c$ axis (color symbols). Solid lines
 and shaded areas are simulations using a single-mode cross
 section function and parameters obtained in the analysis of
 const-$q_\parallel$ scans, as described in the text. }
\end{figure}

\subsubsection{Constant-$E$ scans}

As mentioned in the introduction, theory predicts that the
spectrum at $H>H_c$ in the \emph{ideal} AS geometry should lose its
single-mode character. The sharp magnon excitations are expected
to be replaced by a diffuse continuum of states with a lower bound
following the magnon dispersion curve. Even in this case, the
continuum is singular at the lower bound and may be difficult to
distinguish from a single-mode excitation smeared effects of
experimental resolution.

In search for any deviations from the single-mode picture we
performed constant-$E$ scans at 1.8~meV
(Fig.~\ref{exdata3}) and 1.2~meV (not shown) at $H=2$~T$<H_c$,
$H=4$~T$\approx H_c$ and $H=6$~T$>H_c$. These particular energies
were selected to avoid both gap energies at all three field
values. A constant background was assumed for each scan. At each
field, a single-mode profile was simulated using the model
cross-section function described above, with parameter values
obtained in the analysis of const-$q_\parallel$ data. We
found that, to within experimental error, the single-mode model
(solid lines in Fig.~\ref{exdata3}) reproduces all measured
const-$E$ scans very well, below, at and above $H_c$. The observed
variation of scan shape is due to changes in the gap
energies of the two lower modes, shown as shaded areas in
Fig.~\ref{exdata3}. No features beyond those given by the
single-mode approximation could be detected with the resolution of
the present study.

\section{Anisotropic gapped quasi-1D spin system in strong magnetic field: theory}
\label{sec:theory}

We now turn to developing a theoretical model of the high-field
state. Our goal is a semi-quantitative effective field theoretical
description, backed by a simple physical picture, yet capable of
consistently reproducing all the available experimental data on
NDMAP. The latter implies that the model should account for both
neutron and ESR measurements,\cite{Honda99,Hagiwara+03}
 work both above and below
the critical field, and apply in the case of arbitrary field
orientation.

\subsection{A single anisotropic Haldane chain in a field}

\subsubsection{Existing models}

In the early 90s, several phenomenological field-theoretical
descriptions of the high-field regime in the anisotropic Haldane
chain were proposed.\cite{Affleck90-2,Affleck91,Tsvelik90,Mitra94}
Affleck \cite{Affleck90-2,Affleck91} proposed a theory based on
coarse-graining the $O(3)$ nonlinear sigma model (NLSM).\cite{Haldane83,Haldane83-2}
Technically coarse-graining leads to relaxing the unit vector
constraint of the NLSM, so that one has a theory of unconstrained
real vector bosonic field $\bm{\varphi}$. The $\varphi^{4}$-type
interaction was added to ensure stability in the high-field
regime, and the anisotropy was introduced just by assuming  three
different masses $\Delta_{\alpha}$ for the three field components,
so that the resulting Lagrangian had the form
\begin{eqnarray}
\label{lag:affleck}
\mathcal{L}&=&\frac{1}{2v}  \Big\{
(\partial_{t}\bm{\varphi}
+ \bm{H}\times\bm{\varphi})^{2} -v^{2}(\partial_{x}\bm{\varphi})^{2}
\nonumber\\
&-&\sum_{\alpha}\Delta_{\alpha}\varphi_{\alpha}^{2}
-\lambda(\bm{\varphi}^{2})^{2}
\Big\}.
\end{eqnarray}
For $H>H_{c}$
the ground state acquires a nonzero staggered magnetization
$\bm{L}=\langle\bm{\varphi}\rangle$ and a uniform magnetization
$\bm{M}\propto \langle\bm{H}\times \bm{\varphi}\rangle$. This
model captures the basic physics involved, but is known to suffer from
several drawbacks. Because of the too simplistic way
of introducing the anisotropy, the predicted values of the critical
field ($H_{c}^{(\alpha)}=\Delta_{\alpha}$ for the field $\bm{H}$
directed along one of the symmetry axes $\bm{e}_{\alpha}$)  disagree
with the results of perturbative treatment
\cite{Golinelli92,Golinelli93-2,Regnault93}
as well as with the experimental data on the behavior of gaps as
functions of the applied field in NENP
\cite{Regnault94} and NDMAP.\cite{Zheludev+03}

Tsvelik \cite{Tsvelik90} proposed a different theory which stems from
the integrable Takhtajan-Babujian model of a $S=1$ chain and involves
three Majorana fields with masses $\Delta_{\alpha}$.  The theory
yields the critical field value
$H_{c}^{(\alpha)}=\sqrt{\Delta_{\beta}\Delta_{\gamma}}$ which
coincides with the perturbative formulas of
\cite{Golinelli92,Golinelli93-2,Regnault93}.  This model was rather successful for the
description of the field dependencies of the gaps below $H_{c}$ in
NENP,\cite{Regnault94} and is expected to yield a correct critical
behavior at $H\to H_{c}$.  However, when the high-field neutron data
on NDMAP in the $H\parallel a$ geometry became available,
\cite{Zheludev+03}
 it turned out that
Tsvelik's theory, apart from overestimating the $H_{c}$ value
considerably, predicts no change of slope for the two upper magnon
modes at $H=H_{c}$, in complete disagreement with the experimental
data. One may conclude that though this model correctly describes
the behavior of the low-energy degrees of freedom near the
critical field, but fails to describe the behavior of high-energy
modes above $H_{c}$.

Mitra and
Halperin \cite{Mitra94} have modified Affleck's bosonic Lagrangian
in order to reproduce Tsvelik's results for the gaps.
It turns out that
changing the first term in (\ref{lag:affleck}) to
\begin{equation}
\label{lag:mitra}
\frac{1}{2v}\sum_{\alpha}\big\{\partial_{t}\varphi_{\alpha}
+\sum_{\beta\gamma} (\Delta_{\gamma}/\Delta_{\alpha})^{1/2}
\epsilon_{\alpha\beta\gamma} H_{\beta} \varphi_{\gamma}
\big\}^{2}
\end{equation}
exactly reproduces the results of Ref.\ \onlinecite{Tsvelik90} for the
field dependencies of the gaps
\emph{below} $H_c$.
Above $H_c$, the predicted
field behavior of the gaps is different from that of Ref.\
\onlinecite{Tsvelik90}, and is in a reasonable qualitative agreement
with the experimental data on NDMAP in the $H\parallel a$
geometry.\cite{Zheludev+03}
However, apart from the fact that the reasons for the postulated
modification (\ref{lag:mitra})
remain unclear, the theory still has one fundamental
flaw: it predicts that the staggered moment at $H>H_{c}$ is
directed along \emph{the magnetic hard axis $c$} for $H\parallel a$ and
along the intermediate axis $b$ for $H\parallel c$. This result is
not only counter-intuitive but also
contradicts to the diffraction experiments on NDMAP
\cite{Chen01} which show that the ordered moment, in complete analogy
to the classical picture for an ordered antiferromagnet, always lies in the
easy plane along the most easy axis perpendicular to the field,
i.e., along the $b$ axis for $H\parallel a$  and along the $a$ axis in the
$H\parallel c$ case.

\subsubsection{An improved model}

We see that none of the previously known models provides a consistent description of
the experimental data. We use a different, more general approach, based on the model
proposed in Ref.~\onlinecite{Kolezhuk96} for dimerized $S=1/2$ chains and $S=1/2$
ladders, known to be in the same universality class as $S=1$ Haldane chains. This
model was recently applied with great success to the description of the INS data for
NDMAP in the $H\parallel a$ geometry,\cite{Zheludev+03} and to ESR
experiments in both geometries.\cite{Hagiwara+03}

We first illustrate the general features of the theory on the
example of the  alternated $S=\frac{1}{2}$ chain consisting of
weakly coupled anisotropic dimers, described by the Hamiltonian
\begin{equation}
\label{ham:anis}
{\mathcal H}=\sum_{n\alpha}
J_{\alpha}S_{2n-1}^{\alpha}S_{2n}^{\alpha} +\sum_{n} \{ J'
(\vec{S}_{2n}\cdot\vec{S}_{2n+1})
-\vec{H}\cdot\vec{S}_{n} \},
\end{equation}
where $0<J' \ll J$. Throughout the rest of this section,
it is implied that the magnetic field is measured in energy units,
i.e., $H\mapsto g\mu_{B}H$ unless explicitly stated otherwise.
For the derivation of the effective field theory it is convenient to use the dimer
coherent states \cite{Kolezhuk96}
\begin{equation}
|\vec{A},\vec{B}\rangle=(1-A^2-B^2)^{1/2}
|s\rangle+\sum_j (A_{j}+iB_{j})|t_j\rangle,
\label{wf}
\end{equation}
where the singlet state $|s\rangle$ and
three triplet states $|t_j\rangle$ are given by
\cite{SachdevBhatt90}
\begin{eqnarray*}
|s\rangle=\frac{1}{\sqrt2}
\big(|\uparrow\downarrow\rangle-|\downarrow\uparrow\rangle\big),
&\quad& |t_z\rangle=\frac{1}{\sqrt2}
\big(|\uparrow\downarrow\rangle+
|\downarrow\uparrow\rangle\big),\\
|t_x\rangle=-\frac{1}{\sqrt2}
\big(|\uparrow\uparrow\rangle-|\downarrow\downarrow\rangle\big),
&\quad& |t_y\rangle=\frac{i}{\sqrt2}
\big(|\uparrow\uparrow\rangle+|\downarrow\downarrow\rangle\big),
\end{eqnarray*}
and $\vec{A}$, $\vec{B}$ are real vectors which are in a simple manner
connected with the magnetization
$\vec{M}=\langle\vec{S}_1+\vec{S}_2\rangle$
and sublattice magnetization
$\vec{L}=\langle\vec{S}_1-\vec{S}_2\rangle$ of the
spin dimer:
\begin{equation}
\vec{M}=2(\vec{A}\times\vec{B})\;,\quad
\vec{L}=2(1-A^2-B^2)^{1/2}\vec{A}\;.
\label{ML}
\end{equation}
The configuration space is the inner domain of the unit sphere
$\vec{A}^2+\vec{B}^2\leq1$ in $R^6$, with additional
identification of the opposite points on the sphere, and the
measure is defined as $6\pi^{-3} d\vec{A}\;d\vec{B}$.

We will assume that we are not too far above the critical field, so that the
magnitude of the triplet components is small, $A,B\ll1$.
Then the effective Lagrangian density in the continuum limit takes the following form:
\begin{eqnarray}
\label{Leff-anis}
{\mathcal L}&=& -2\hbar \vec{B}\cdot\partial_{t}\vec{A} -\frac{1}{2}J'
\ell^{2}(\partial_{x}\vec{A})^{2} -\sum_{i}\{
m_{i}A_{i}^{2}
+\widetilde{m}_{i}B_{i}^{2} \}\nonumber\\
&+& 2\vec{H}\cdot(\vec{A}\times\vec{B})-U_{4}(\vec{A},\vec{B}),
\end{eqnarray}
where $m_{i}=\widetilde m_{i}-J'$, $\ell$ is the lattice
constant, and $\widetilde
m_{i}=\frac{1}{4}|\epsilon_{ijn}|
(J_{j}+J_{n})$. The fourth-order term
\begin{equation}
\label{U4AB}
V_{4}(\vec{A},\vec{B})
=\lambda(\vec{A}^{2})^{2}
+\lambda_{1}(\vec{A}^{2}\vec{B}^{2})+\lambda_{2}(\vec{A}\cdot\vec{B})^{2},
\end{equation}
where $\lambda=J'$, $\lambda_{1}=2J'$, $\lambda_{2}=-J'$ in the
present case.  The spatial derivatives of $\vec{B}$ are omitted in
(\ref{Leff-anis}) because they appear only in terms which are of the
fourth order in $\vec{A}$, $\vec{B}$. Generally, we can assume that
spatial derivatives are small (small wave vectors), but we shall not
assume that the time derivatives (frequencies) are small since we are
going to describe high-frequency modes as well.

The vector
$\vec{B}$ can be integrated out, and under the assumption $A\ll 1$ it can be
expressed through $\vec{A}$ as follows:
\begin{eqnarray}
\label{B}
&&\vec{B}=\widehat{Q}\vec{F},\quad \vec{F}= -\hbar\partial_{t}\vec{A}
+(\vec{H}\times\vec{A}) \nonumber\\
&& Q_{ij}={\delta_{ij}\over \widetilde{m}_{i} }
-\lambda_{2}{A_{i}A_{j}\over \widetilde{m}_{i}\widetilde{m}_{j}} .
\end{eqnarray}
After substituting this expression back into (\ref{Leff-anis}) one obtains the
effective Lagrangian depending on $\vec{A}$ only:
\begin{eqnarray}
\label{Leff-anis-A}
{\mathcal L}&=& {1\over \widetilde{m}_{i}}
\hbar^{2}(\partial_{t} A_{i})^{2} -v_{i}^{2}\ell^{2}(\partial_{x}A_{i})^{2}\Big\} \\
&-&2{\hbar\over \widetilde{m}_{i}} (\vec{H}\times\vec{A})_{i}
\partial_{t} A_{i} -U_{2}(\vec{A})
  -U_{4}(\vec{A},\partial_{t}\vec{A}),\nonumber
\end{eqnarray}
where $v_{i}=\sqrt{J'\widetilde{m}_{i}/2}$ are the characteristic velocities in
energy units, and
the quadratic and quartic parts of the potential are given by
\begin{eqnarray}
\label{U2-A}
&& U_{2}(\vec{A})=
m_{i}A_{i}^{2}-{1\over\widetilde{m}_{i}}(\vec{H}\times\vec{A})_{i}^{2},\\
&& U_{4}(\vec{A},\partial_{t}\vec{A})= \lambda (\vec{A}^{2})^{2} +
\lambda_{1} \vec{A}^{2} {1\over \widetilde{m}_{i}^{2}}F_{i}^{2}+
\lambda_{2}{A_{i}A_{j}\over \widetilde{m}_{i}\widetilde{m}_{j}} F_{i}F_{j}\nonumber
\end{eqnarray}
Note that the cubic in $\vec{A}$ term in (\ref{B}) must be kept since
it contributes to the $U_{4}$ potential.

Having in mind that the alternated $S={1\over2}$ chain and the Haldane
chain belong to the same universality class, one may now try to apply
this model in the form (\ref{Leff-anis-A}-\ref{U2-A}) to the Haldane
chain system such as NDMAP, treating the velocities $v_{i}$ and
interaction constants $m_{i}$, $\widetilde{m}_{i}$, $\lambda$, and
$\lambda_{1,2}$ as phenomenological parameters.

One can show that the Lagrangian (\ref{Leff-anis-A}) contains theories of Affleck
\cite{Affleck90-2,Affleck91} and Mitra and Halperin \cite{Mitra94} as particular
cases. Indeed, restricting the interaction to the simplified form with
$\lambda_{1,2}=0$ and assuming isotropic velocities $v_{i}=v$, one can see that the
Affleck's Lagrangian (\ref{lag:affleck}) corresponds to the isotropic
$\vec{B}$-stiffness $\widetilde{m}_{i}=\widetilde{m}$, while another choice
$\widetilde{m}_{i}=m_{i}$ yields the modification (\ref{lag:mitra}).

For illustration, let us assume  that $\vec{H}\parallel z$. Then the
quadratic part of the potential takes the form
\begin{equation}
\label{U2-z}
U_{2}=(m_{x}-{H^{2}\over \widetilde{m}_{y}})A_{x}^{2}
+(m_{y}-{H^{2}\over \widetilde{m}_{x}})A_{y}^{2}
+m_{z}A_{z}^{2},
\end{equation}
and the critical field is obviously
$H_{c}=\min\{(m_{x}\widetilde{m}_{y})^{1/2},(m_{y}\widetilde{m}_{x})^{1/2} \}$.
At zero field the three triplet gaps are given by
$\Delta_{i}=(m_{i}\widetilde{m}_{i})^{1/2}$.
Below
$H_{c}$ the energy gap for the mode polarized along the field stays constant,
$\hbar\omega_{z}=\Delta_{z}$, while the gaps for the other two modes are
given by
\begin{eqnarray}
\label{Exy}
(\hbar\omega_{xy}^{\pm})^{2}&=&{1\over2}(\Delta_{x}^{2}+\Delta_{y}^{2}) +H^{2} \\
&\pm&
\Big[
(\Delta_{x}^{2}-\Delta_{y}^{2})^{2}
  +H^{2}(m_{x}+m_{y})(\widetilde{m}_{x} +\widetilde{m}_{y})
\Big]^{1/2}.\nonumber
\end{eqnarray}
Below $H_{c}$ the mode energies do not depend on the interaction
constants $\lambda_{i}$.

It is easy to see that in the special case $m_{i}=\widetilde{m}_{i}$, the above
expression transforms into
\begin{equation}
\label{Exy-tsv}
\hbar\omega_{xy}^{\pm}={1\over2}(\Delta_{x}+\Delta_{y})\pm \Big[
  {1\over4}(\Delta_{x}-\Delta_{y})^{2}+H^{2}\Big]^{1/2},
\end{equation}
which exactly coincides with the formulas obtained in the approach of Tsvelik,
\cite{Tsvelik90} and also with the perturbative formulas of
Ref.~\onlinecite{Golinelli92,Golinelli93-2,Regnault93} and with the results of modified
bosonic theory of Mitra and Halperin. \cite{Mitra94} The
peculiarity of this
special choice of parameters is that above
$H_{c}=(m_{x}m_{y})^{1/2}$
the most negative eigenvalue of the quadratic form
(\ref{U2-z}) corresponds not to the component of $\vec{A}$ along the
easy axis in the $(xy)$ plane, as one would intuitively expect, but to
the component along the harder axis. For instance, if $x$ is the easy
axis, $m_{x}<m_{y}$, then the most negative coefficient will be that
at $A_{y}$. This leads to the above-mentioned problem with
counter-intuitive direction of the ordered moment in the
theory of Mitra and Halperin.

Generally, at $H>H_{c}$ one has to find the minimum of the static part of the
potential
and linearize the theory around the new static solution
$\vec{A}=\vec{A}^{(0)}$. The equations for $\vec{A}^{(0)}$ have the form
\begin{equation}
\label{eq-A0}
\sum_{j}\Omega_{\beta j}A_{j} +\sum_{imn}\Lambda_{\beta i,mn} A_{i} A_{m}A_{n}=0,
\end{equation}
where the matrices $\mathsf{\Omega}$, $\mathsf{\Lambda}$ are defined as
\begin{eqnarray}
\label{mat1}
&& \Omega_{ij}=m_{i}\delta_{ij}
-\sum_{kln}\epsilon_{ikn}\epsilon_{jln}\frac{1}{\widetilde{m}_{n}}
H_{k}H_{l}, \nonumber\\
&& \Lambda_{ij,mn}=\Gamma_{ij,mn}+\Gamma_{mn,ij},\\
&& \Gamma_{ij,mn}=\lambda \delta_{ij}\delta_{mn}
+(\lambda_{1}\delta_{ij} +\lambda_{2})\sum_{kl}
\epsilon_{ikn}\epsilon_{jlm}\frac{ H_{k}H_{l} }{
\widetilde{m}_{i}\widetilde{m}_{j} }.
\nonumber
\end{eqnarray}
The magnon energies $\hbar\omega$ as functions of the field $H$ and of the
longitudinal (with respect to the chain direction)
wave vector
$q$ can be found as three real roots
of the secular equation
\begin{equation}
\label{secular}
\det( \mathsf{M} -(\hbar\omega)^{2}\mathsf{G}
-i\hbar\omega\mathsf{C})=0,
\end{equation}
where symmetric matrices $\mathsf{M}$ and $\mathsf{G}$ are
given by
\begin{eqnarray}
\label{mat2}
 M_{ij}&=&\Omega_{ij} +
\frac{(v_{i} q \ell )^{2}}{\widetilde{m}_{i}}\delta_{ij} \nonumber\\
&+& \sum_{mn} A_{m}^{(0)} A_{n}^{(0)}
(\Lambda_{ij,mn}+\Lambda_{im,jn}+\Lambda_{in,mj}),\nonumber\\
 G_{ij}&=&\left\{ \frac{1}{\widetilde{m}_{i}} -\lambda_{1}\Big(
\frac{ A_{i}^{(0)}}{\widetilde{m}_{i}}\Big)^{2} \right\} \delta_{ij}
-\lambda_{2}\frac{ A_{i}^{(0)} A_{j}^{(0)} }{
\widetilde{m}_{i}\widetilde{m}_{j}},
\end{eqnarray}
and the antisymmetric matrix
$\mathsf{C}=\mathsf{R}-\mathsf{R}^{T}$ is determined by
\begin{eqnarray}
\label{mat3}
R_{ij}&=&\left\{ \frac{1}{\widetilde{m}_{i}} -\lambda_{1}\Big(
\frac{ A_{i}^{(0)}}{\widetilde{m}_{i}}\Big)^{2} \right\}
\sum_{l}\epsilon_{ijl}H_{l} \nonumber\\
&+&\lambda_{2} \frac { A_{i}^{(0)}}{\widetilde{m}_{i}} \sum_{ln}
\Big\{
\frac{1}{\widetilde{m}_{j}} -
\frac{1}{\widetilde{m}_{n}}\Big\} \epsilon_{jln} A_{n}^{(0)}
H_{l}.
\end{eqnarray}
Here in case of NDMAP the longitudinal wave vector $q$
must be understood as counted from the 1D  Bragg point,
$q\ell\mapsto q_{\parallel}c-\pi$.

\subsection{Inter-chain interactions}

Up to now, we have treated the problem as purely one-dimensional.
In NDMAP, however, antiferromagnetic interchain interactions along
the crystallographic $b$ direction lead to an observable
transverse dispersion with the bandwidth of about
$0.1$~meV.\cite{Zheludev2001} This value is small compared to the
magnon bandwidth along the chain axis ($\simeq 7$~meV), but
constitutes approximately 20\% of the lowest magnon gap, so that
transverse interactions have to be taken into account if one aims
at a quantitative description.

 It is straightforward to incorporate this effect into our
formalism.
Additional coupling of the form
\begin{equation}
\label{transverse}
J_{\perp}\sum_{j}(\vec{A}_{j}\cdot\vec{A}_{j+1}),
\end{equation}
where $j$ labels chains along the transverse $b$-direction,
amounts to a renormalization of  the $\vec{A}$-stiffnesses
$m_{\alpha}\mapsto m_{\alpha}+2J_{\perp}\cos (q_{\perp}b)$. The
minimum of magnon energies is reached at the 3D AF zone center, in
our case at $q_{\parallel}c=q_{\perp}b=\pi$. It is convenient to
\emph{redefine} the stiffness $m_{\alpha}$ as its value at the
zone center $m_{\alpha}(q_{\perp}=\pi/b)$, so that in the formulas
(\ref{eq-A0})-(\ref{mat3}) one just has to make the substitution
\[
m_{\alpha}\mapsto  m_{\alpha}+2J_{\perp}(1+\cos
q_{\perp}b).
\]
 Secular equations (\ref{secular}) then yield magnon energies for an
arbitrary transverse wave vector transfer.

\begin{figure}
\includegraphics[width=3.5in]{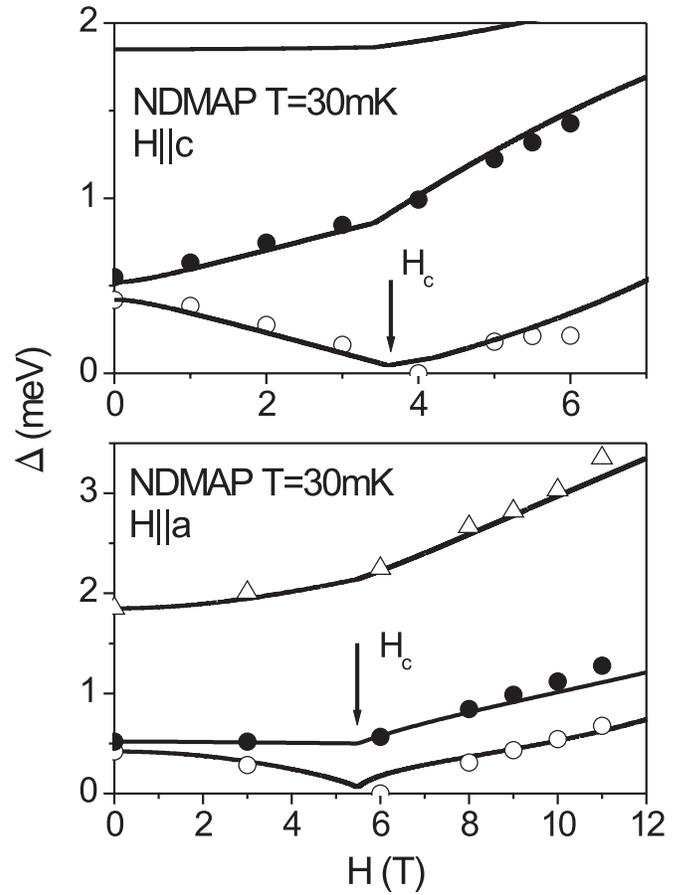}
 \caption{\label{gapvsH} Symbols: measured field dependence of the
 energy gaps in NDMAP for a magnetic field applied along the
 crystallographic $c$-axis (top, this work) and $a$-axis (bottom,
 Ref.~\protect\onlinecite{Zheludev+03}). Lines: a global fit to
 the experimental data based on the Ginsburg-Landau type model
 outlined in Ref.~\protect\onlinecite{Zheludev+03} and described
 in detail in the text. Note the different scales on the top and
 bottom figures.}
\end{figure}

\subsection{Comparison with experiment}

The main advantage of the described model is that it can
consistently reproduce all the experimental data currently
available for NDMAP. At a first glance, it may appear
over-parameterized, with 9 separate phenomenological constants:
$m_{\alpha}$, $\widetilde{m}_{\alpha}$, $\lambda$, $\lambda_1$ and
$\lambda_2$. However, all these parameters are relevant and can be
almost uniquely determined from the measured field dependencies of
the gap energies.  Indeed, as discussed above, the independently
measured zero-field gaps $\Delta_\alpha$ fix three relations
$\Delta_\alpha^2=m_\alpha \widetilde{m}_{\alpha}$. The value of
the critical field $H_c^{(\alpha)}$ for a field applied along the
principal anisotropy axis $\alpha$  determines another three
relations between parameters, namely $(H_c^{(\alpha)})^2=
\min(m_\beta\widetilde{m}_{\gamma},
m_\gamma\widetilde{m}_{\beta})$. One can show that of those six
equations for six stiffness constants $m_{\alpha}$ and
$\widetilde{m}_{\alpha}$ only five are independent, so that all
stiffness constants can be expressed through one of them (we have
chosen $m_{x}$ for this role). Finally, the interaction parameters
$\lambda$, $\lambda_{1,2}$  control the behavior of the gap
energies above the critical fields. In fact, the gaps depend only
on the relative interaction strengths $(\lambda_{1}/\lambda)$,
$(\lambda_{2}/\lambda)$, so that the scale of $\lambda$ does not
influence the expressions for the gaps and can be set deliberately
(we have put $\lambda=1$). Thus, the knowledge of the $H=0$ gaps
and of critical fields helps to fix five parameters, and one
parameter turns out to be irrelevant, so that one is left with
only three parameters to fit the $\Delta(H)$ curves.

In analyzing the measured field dependencies of the gap energies
in NDMAP, one has to keep in mind that both for the experiments
described here, and for those reported in
Ref.~\onlinecite{Zheludev+03} for $H\parallel a$, the transverse
wave vector $q_{\perp}=2\pi k/b$ is not constant, but varies as a
function of energy transfer $\hbar \omega$. For $H\parallel c$,
$q_\perp$ is dictated by the geometry of the horizontal-field
magnet, $q_{\perp}b/(2\pi) = 0.42\,{\rm meV}^{-1} \times
(\hbar\omega)$. In the $H\parallel a$ experiment of
Ref.~\onlinecite{Zheludev+03} $q_\perp$ was chosen to optimize
wave vector resolution along the chains when using a horizontally
focusing analyzer, $q_\perp b/(2\pi)=1.3+0.24\,{\rm meV}^{-1}
\times (\hbar\omega)$. In both cases the presence of inter-chain
interactions must be taken into account explicitly, as discussed
in the previous subsection.  The value of $4J_{\perp}$ was chosen
to match the transverse dispersion bandwidth of $0.1$~meV observed
in NDMAP \cite{Zheludev2001} at $H=0$: $J_\perp=0.025$~meV.

A global fit to the data of both experiments is shown in
Fig.~\ref{gapvsH}. The coordinate axes were chosen along the
\emph{local} anisotropy axes for each Ni$^{2+}$ ion: $y$ is
parallel to the $b$ axis, $z$ is in the $(a,c)$ plane and forms an
angle of 16$^{\circ}$ with the $c$ axis and $x$ completes the
orthogonal set. The final parameters obtained in the fit are:
$m_x=0.50$, $m_y=0.71$, $m_z= 4.76$, $\widetilde{m}_{x}=0.35$,
$\widetilde{m}_{y}=0.38$, $\widetilde{m}_{z}=0.71$,
$\lambda_1/\lambda=0.17$ and $\lambda_2/\lambda=-0.17$. The solid
lines in Fig.~\ref{gapvsH} never quite reach zero at $H_c$, since
they actually represent excitation energies at $\mathbf{q}_\|=\pi$
and transverse wave vector transfers matching those probed in the
corresponding experiments. At the critical field these energies
are non-zero due to non-zero dispersion perpendicular to the chain
axis (see above discussion). Note that the fitted critical fields
$H_c^{(a)}=5.5$~T and $H_c^{(c)}=3.4$~T are some 15\% smaller than
observed in our neutron scattering experiments. Incidentally,
these critical fields are in excellent agreement with those found
in heat capacity\cite{Honda98} ESR measurements.\cite{Hagiwara+03}
The parameter values obtained in the present neutron study also
agree nicely with those determined in the analysis of ESR
resonance frequencies. Finally, we have verified and would like to
stress that the obtained parameter values yield \emph{correct}
directions of the ordered staggered moment at $H>H_c$ for  both
$H\parallel c$  and $H\parallel a$ experimental geometries:
namely, the order parameter $\vec{A} $ is directed along the $b$
and $a$ axes in the $H\parallel a$ and $H\parallel c$ cases,
respectively.

\section{Concluding remarks}
\label{sec:summary}

Even for $H\parallel c$ NDMAP shows no signs of Luttinger spin
liquid behavior, such as incommensurate correlations or breakdown
of the single-particle spectrum. On the contrary, the system was
shown to be antiferromagnetically {\it ordered} (a ``spin solid''
state) at high fields with an appreciable sublattice
magnetization. This is accompanied by a re-opening of the gap at
$H>H_c$.  Overall, the observed field dependence of the excitation
spectrum is qualitatively similar to that previously seen for
$H\perp c$.\cite{Zheludev+03}  The reasons for the Luttinger spin
liquid regime being unobservable are quite clear: the idealized
axially symmetric geometry can not be realized in NDMAP. This, to
our opinion, may be mainly attributed to the effects of the
16$^{\circ}$ canting of the principal anisotropy axes relative to
the field direction, which define the strongest explicit breaking
of the axial symmetry, although in-plane anisotropy and
inter-chain interactions play a significant role as well.  Perhaps
in very strong external fields, when the Zeeman energy becomes
large compared to any anisotropy and 3D effects, certain features
of the Luttinger spin liquid may become accessible. In particular,
it has been recently argued\cite{Wang2003} that incommensurate
correlation should emerge even in the axially asymmetric geometry,
above a certain second critical field $H_c^{\ast}$. Whether or not
such an experiment is technically feasible for NDMAP is currently
unclear.

In summary, the main impact of the present $H\parallel c$ measurements is to provide
crucial quantitative data needed to evaluate the veracity of the various
field-theoretical descriptions of magnetized anisotropic Haldane spin chains. Our
conclusion is that of several existing theories, only the newly proposed model is
robust enough to reproduce all the features observed experimentally for different
field orientations.

\acknowledgments We would like to acknowledge Collin L. Broholm
who played a key role in earlier experiments on NDMAP and was
intellectually involved in all the studies described in this work.
The expertise of Peter Smeibidl was crucial in setting up and
maintaining the high-field and low temperature sample environment
during experiments at HMI.  We would also like to thank
F.~E{\ss}ler, A.~Tsvelik, and I.~Zaliznyak for enlightening
discussions. Work at ORNL and BNL was carried out under DOE
Contracts No. DE-AC05-00OR22725 and DE-AC02-98CH10886,
respectively. Work at JHU was supported by the NSF through
DMR-0074571. Experiments at NIST were supported by the NSF through
DMR-0086210 and DMR-9986442. The high-field magnet was funded by
NSF through DMR-9704257. Work at RIKEN was supported in part by a
Grant-in-Aid for Scientific Research from the Japan Society for
the Promotion of Science. Work at ITP Hannover and IMAG Kiev was
partly supported by the grant I/75895 from Volkswagen-Stiftung.


\end{document}